\documentclass[oneside,a4paper,11pt,shownumbers]{article}
\usepackage{latexsym}
\usepackage{euscript}
\usepackage{epsfig,amsmath,amssymb}

\def\be{\begin{equation}}
\def\ee{\end{equation}}
\def\bea{\begin{eqnarray}}
\def\eea{\end{eqnarray}}

\long\def\symbolfootnote[#1]#2{\begingroup%
\def\thefootnote{\fnsymbol{footnote}}\footnote[#1]{#2}\endgroup} 

 \large\normalsize

\begin{document}

\begin{center}

{\Large \bf Angularly excited and interacting boson stars and $Q$-balls}

\vspace*{7mm} {Yves Brihaye $^{a}$
\symbolfootnote[1]{E-mail: yves.brihaye@umh.ac.be}
and  Betti Hartmann $^{b}$
\symbolfootnote[2]{E-mail:b.hartmann@jacobs-university.de}}
\vspace*{.25cm}

${}^{a)}${\it Facult\'e des Sciences, Universit\'e de Mons-Hainaut, 7000 Mons, Belgium}\\
${}^{b)}${\it School of Engineering and Science, Jacobs University Bremen, 28759 Bremen, Germany}\\

\vspace*{.3cm}
\end{center}

\begin{abstract}
We study angularly excited as well as interacting non-topological solitons, so-called $Q$-balls and
their gravitating counterparts, so-called boson stars in $3+1$ dimensions.
$Q$-balls and boson stars carry a non-vanishing Noether charge 
and arise as solutions of complex scalar field models in a flat space-time
background and coupled minimally to gravity, respectively.

We present examples of interacting $Q$-balls that arise due to angular
excitations, which are closely related to the spherical harmonics.
We also construct explicit examples of rotating boson stars that interact with non-rotating boson stars. We observe that rotating boson stars
tend to absorb the non-rotating ones for increasing, but reasonably small gravitational coupling.
This is a new phenomenon as compared to the flat space-time limit and is related to the negative contribution of
the rotation term to the energy density of the solutions.
In addition, our results indicate that a system of a rotating and non-rotating boson star can become
unstable if the direct interaction term in the potential is large enough.
This instability is related to the appearance of ergoregions.

\end{abstract}

\section{Introduction}
Solitons play an important role in many areas of physics. As classical solutions of non-linear field theories, they
are localised structures with finite energy, which are globally regular.
In general, one can distinguish topological and non-topological solitons.
While topological solitons \cite{ms} possess a conserved quantity, the topological charge, that stems (in most
cases) from the spontaneous symmetry breaking of the theory, non-topological solitons \cite{fls,lp} have a conserved Noether
charge that results from a symmetry of the Lagrangian. The standard example of  non-topological solitons
are $Q$-balls \cite{coleman}, which are solutions of theories with self-interacting complex scalar fields. These objects are stationary with an explicitely
time-dependent phase. The conserved Noether charge $Q$
is then related to the global phase invariance of the theory and is a function of
the frequency. $Q$ can e.g. be interpreted as particle number \cite{fls}. 

While in standard scalar
field theories, it was shown
that a non-renormalisable $\Phi^6$-potential is necessary \cite{vw}, supersymmetric extensions of the
Standard Model (SM)  also possess $Q$-ball solutions \cite{kusenko}. In the latter case, several scalar fields
interact via complicated potentials. It was shown that cubic interaction terms that result from
Yukawa couplings in the superpotential and supersymmetry breaking terms lead to the existence of $Q$-balls
with non-vanishing baryon or lepton number or electric charge. These supersymmetric
$Q$-balls have been considered recently as possible candidates for baryonic dark matter 
\cite{dm} and their astrophysical implications have been discussed \cite{implications}.
In \cite{cr}, these objects have been constructed numerically using 
the exact form of the supersymmetric potential.

$Q$-ball solutions in $3+1$ dimensions have been studied in detail in 
\cite{vw,kk1,kk2}. It was realised
that next to non-spinning $Q$-balls, which are spherically symmetric, spinning solutions
exist. These are axially symmetric with energy density of toroidal shape
and angular momentum $J=kQ$, where $Q$ is the Noether charge of the solution
and $k\in \mathbb{Z}$ corresponds to the winding around the $z$-axis. 
Approximated  solutions of the non-linear partial differential equations
were constructed in \cite{vw} by means of a truncated series in the spherical harmonics to describe
the angular part of the solutions. 
The full  partial differential equation was solved numerically in \cite{kk1,kk2,bh}
(for a short review see \cite{radu}). 
It was also realised in \cite{vw} that in each $k$-sector, parity-even ($P=+1$)
and parity-odd ($P=-1$) solutions exist. Parity-even and parity-odd 
refers to the fact that
the solution is symmetric and anti-symmetric, respectively with respect
to a reflection through the $x$-$y$-plane, i.e. under $\theta\rightarrow \pi-\theta$.

These two types of solutions are
closely related to the fact that the angular part of the solutions
constructed in \cite{vw,kk1,kk2}
is connected to the spherical harmonic $Y_0^0(\theta,\varphi)$ for the spherically symmetric $Q$-ball,
to the spherical harmonic $Y_1^{1}(\theta,\varphi)$ for the spinning parity even ($P=+1$) solution
and to the spherical harmonic $Y_2^{1}(\theta,\varphi)$  for the parity  odd ($P=-1$) solution, respectively. 
Radially excited solutions of the spherically symmetric, non-spinning solution were also obtained.
These solutions are still spherically symmetric but the scalar field develops one or several nodes for $r\in ]0,\infty[$. 
In relation to the apparent connection of the angular part of the known solutions to the spherical harmonics,
``$\theta$-angular excitations'' of the $Q$-balls 
corresponding to the spherical harmonics  $Y_l^k(\theta,\varphi)$, $-l \leq k \leq l$ have been constructed
explicitely for some values of $k$ and $l$ in \cite{bh}.
These excited solutions could play a role in the formation of $Q$-balls in the early universe
since it is believed that $Q$-balls forming due to condensate fragmentation at the end of inflation first appear in an excited state and only then settle down to the ground state \cite{nonspherical}. The fact that these newly formed $Q$-balls are excited, i.e.
in general not spherically symmetric could, on the other hand,  be a source of gravitational waves \cite{km2}.

The interaction of two $Q$-balls has also been studied in \cite{bh}.
It was found that the lower bound on the frequencies $\omega_i$, $i=1,2$
is increasing for increasing interaction coupling. 
Explicit examples of a rotating $Q$-ball interacting with a non-rotating
$Q$-ball have been presented.

Complex scalar field models coupled to gravity exhibit
a new type of solution, so-called ``boson stars'' \cite{misch,flp,jetzler}.
In \cite{kk1,kk2} boson stars have
been considered that have flat space-time limits in the form of
$Q$-balls.

In this paper, we study interacting boson stars with flat limit
solutions in the form of the interacting $Q$-balls that have been studied in
\cite{bh}. The boson stars are interacting via a potential term, but of course also through gravity.

The paper is organised as follows: in Section 2, we give the model, Ansatz and boundary conditions. In Section 3 and 4, we discuss our results for $Q$-balls and their gravitating
counterparts, boson stars, respectively. Section 5 contains our conclusions.

\section{The model}
In the following, we study a scalar field model coupled minimally to gravity in $3+1$ dimensions describing two interacting boson stars. 
The action $S$ reads:
\begin{equation}
 S=\int \sqrt{-g} d^4 x \left( \frac{R}{16\pi G} + {\cal L}_{m}\right)
\end{equation}
where $R$ is the Ricci scalar, $G$ denotes Newton's constant and ${\cal L}_{m}$ denotes
the matter Lagrangian: 
\begin{equation}
\label{lag}
 {\cal L}_{m}=-\frac{1}{2}\partial_{\mu} \Phi_1 \partial^{\mu} \Phi_1^*-
\frac{1}{2}\partial_{\mu} \Phi_2 \partial^{\mu} \Phi_2^* - V(\Phi_1,\Phi_2)
\end{equation}
where both $\Phi_1$ and $\Phi_2$ are complex scalar fields and we choose as signature of the metric
$(-+++)$. The potential reads:
\begin{equation}
V(\Phi_1,\Phi_2)=\sum_{i=1}^2\left(
\kappa_i \vert\Phi_i\vert^6 - \beta_i \vert\Phi_i\vert^4 +
\gamma_i \vert\Phi_i\vert^2 \right)
+\lambda \vert\Phi_1\vert^2 \vert\Phi_2\vert^2
\end{equation} 
where $\kappa_i$, $\beta_i$, $\gamma_i$, $i=1,2$ are the standard potential parameters for each boson star, while $\lambda$ denotes the interaction parameter. The masses of the two bosonic scalar fields are then given by $(m_B^i)^2 = \gamma_i$, $i=1,2$.

Along with
\cite{kk1,kk2,bh}, we choose in the following 
\begin{equation}
\label{parameters}
 \kappa_i=1 \ \ , \ \ \beta_i=2 \ \ , \ \ \gamma_i=1.1 \ \ , \ \ i=1,2 \ .
\end{equation}

In \cite{vw} it was argued that a $\Phi^6$-potential is necessary in order to have classical
$Q$-ball solutions. This is still necessary for the model we have defined here,
since we want $\Phi_1=0$
and $\Phi_2=0$ to be a local minimum of the potential. A pure $\Phi^4$-potential which is bounded from below wouldn't fulfill these criteria. 
 
The matter Lagrangian ${\cal L}_{m}$ (\ref{lag}) is invariant under the two independent global U(1) transformations
\begin{equation}
 \Phi_1 \rightarrow \Phi_1 e^{i\chi_1} \ \ \ , \ \ \ 
\Phi_2 \rightarrow \Phi_2 e^{i\chi_2}   \ .
\end{equation}
As such the total conserved Noether
current $j^{\mu}_{(tot)}$, $\mu=0,1,2,3$, associated to these symmetries is just the sum of
the two individually conserved currents $j^{\mu}_{1}$ and $j^{\mu}_2$ with
\begin{equation}
j^{\mu}_{(tot)}= j^{\mu}_1 +j^{\mu}_2
 = -i \left(\Phi_1^* \partial^{\mu} \Phi_1 - \Phi_1 \partial^{\mu} \Phi_1^*\right)
-i  \left(\Phi_2^* \partial^{\mu} \Phi_2 - \Phi_2 \partial^{\mu} \Phi_2^*\right)\ \ .
\end{equation}
with $j^{\mu}_{1\ ;\mu}=0$, $j^{\mu}_{2 \ ;\mu}=0$ and  $j^{\mu}_{(tot) \ ; \mu}=0$.

The total Noether charge $Q_{(tot)}$ of the system is then the sum of the two individual Noether charges $Q_1$ and $Q_2$:
\begin{equation}
 Q_{(tot)}=Q_1+Q_2= -\int j_1^0 d^3 x  -\int j_2^0 d^3 x
\end{equation}

Finally, the energy-momentum tensor reads:
\begin{equation}
\label{em}
T_{\mu\nu}=\sum_{i=1}^2 \left(\partial_{\mu} \Phi_i \partial_{\nu} \Phi_i^*
+\partial_{\nu} \Phi_i \partial_{\mu} \Phi_i^*\right) -g_{\mu\nu} {\cal L}
\end{equation}

\subsection{Ansatz and Equations}
For the metric the Ansatz in Lewis-Papapetrou form reads \cite{kk1}:
\begin{equation}
 ds^2 = - fdt^2 + \frac{l}{f}\left(g (dr^2 + r^2 d\theta^2) + r^2 \sin^2 \theta (d\varphi + \frac{m}{r} dt)^2 \right)
\end{equation}
where the metric function $f$, $l$, $g$ and $m$ are functions of $r$ and $\theta$ only.
For the scalar fields, the Ansatz reads:
\begin{equation}
\label{ansatz1}
\Phi_i(t,r,\theta,\varphi)=e^{i\omega_i t+ik_i\varphi} \phi_i(r,\theta) \  \ ,
\ i=1,2
\end{equation}
where the $\omega_i$ and the $k_i$ are constants. Since we require $\Phi_i(\varphi)=\Phi_i(\varphi+2\pi)$, $i=1,2$, we have $k_i\in \mathbb{Z}$.
The mass $M$ and total angular momentum $J$ of the solution can be read off from the asymptotic behaviour
of the metric functions \cite{kk1}:
\begin{equation}
 M=\frac{1}{2G} \lim_{r\rightarrow \infty} r^2 \partial_r f  \ \ , \ \
J=\frac{1}{2G} \lim_{r\rightarrow \infty} r^2 m   \ .
\end{equation}
The total angular momentum $J=J_1+J_2$ and the Noether charges $Q_1$ and $Q_2$ of the two boson stars are related by $J=k_1 Q_1 + k_2 Q_2$. Boson stars with $k_i=0$ have thus
vanishing angular momentum. Equally, interacting boson stars with $k_1=-k_2$ and $Q_1=Q_2$ have vanishing
angular momentum. 

The coupled system of partial differential equations is then given by the Einstein
equations
\begin{equation}
\label{einstein}
 G_{\mu\nu}=8\pi G T_{\mu\nu}
\end{equation}
with $T_{\mu\nu}$ given by (\ref{em}) and the Klein-Gordon equations 
\begin{equation}
\label{KG}
 \left(\square + \frac{\partial V}{\partial \vert\Phi_i\vert^2} \right)\Phi_i=0 \ \ , \ \ i=1,2 \ .
\end{equation}
Details about these equations for one complex scalar field can e.g. be found in \cite{kk1}.

\subsection{Boundary conditions}

We require the solutions to be regular at the origin. The appropriate boundary conditions
read:
\begin{equation}
\partial_r f|_{r=0}=0 \ , \ \ \
\partial_r l|_{r=0}=0 \ , \ \ \
g|_{r=0}=1 \ , \ \ \
m|_{r=0}=0 \ , \ \ \
\phi_i| _{r =0}=0 \  \ , \ \ i=1,2.
\label{bc3} \end{equation}
for solutions with $k_i\neq 0$, while for $k_i=0$ solutions, we have 
\begin{equation}
\partial_r f|_{r=0}=0 \ , \ \ \
\partial_r l|_{r=0}=0 \ , \ \ \
g|_{r=0}=1 \ , \ \ \
m|_{r=0}=0 \ , \ \ \
\partial_r\phi_i| _{r =0}=0 \  \ , \ \ i=1,2   \ .
\label{bc3new} \end{equation}
The boundary conditions at infinity result from the requirement of asymptotic
flatness and finite energy solutions: 
\begin{equation}
f|_{r \rightarrow \infty} =1 \ , \ \ \
l|_{r \rightarrow \infty} =1 \ , \ \ \
g|_{r \rightarrow \infty} =1 \ , \ \ \
m|_{r \rightarrow \infty} =0 \ , \ \ \
\phi_i| _{r \rightarrow \infty}=0 \ \ , \ \ i=1,2.
\label{bc4} \end{equation}

For $\theta=0$ the regularity of the solutions on the $z$-axis requires:
\begin{equation}
\partial_{\theta} f|_{\theta=0}=0 \ , \ \ \
\partial_{\theta} l|_{\theta=0}=0 \ , \ \ \
g|_{\theta=0}=1 \ , \ \ \
\partial_{\theta} m |_{\theta=0}=0 \ , \ \ \
\phi_i |_{\theta=0}=0 \ \ , \ \ i=1,2 \ , 
\label{bc5} \end{equation}
for $k_i\neq 0$ solutions, while for $k_i=0$ solutions, we have
\begin{equation}
\partial_{\theta} f|_{\theta=0}=0 \ , \ \ \
\partial_{\theta} l|_{\theta=0}=0 \ , \ \ \
g|_{\theta=0}=1 \ , \ \ \
\partial_{\theta} m |_{\theta=0}=0 \ , \ \ \
\partial_{\theta}\phi_i |_{\theta=0}=0 \ \ , \ \ i=1,2 \ , 
\label{bc5new} \end{equation}

The conditions at $\theta=\pi/2$
are either given by
\begin{equation}
\partial_{\theta} f|_{\theta=\pi/2}=0 \ , \ \ \
\partial_{\theta} l|_{\theta=\pi/2}=0 \ , \ \ \
\partial_{\theta} g|_{\theta=\pi/2}=0 \ , \ \ \
\partial_{\theta} m |_{\theta=\pi/2}=0 \ , \ \ \
\partial_{\theta} \phi_i |_{\theta=\pi/2}=0 \ \ , \ \ i=1,2
\label{bc6} \end{equation}
for even parity solutions,
while for odd parity solutions the conditions for the scalar field functions read:
$\phi_i |_{\theta=\pi/2}=0$, $i=1,2$.

\subsection{Numerical procedure}
In the following, we will solve the system of partial differential equations
(\ref{einstein}) and (\ref{KG}) subject to the appropriate boundary
conditions given in Section 2.2. This has been done using the PDE solver FIDISOL
\cite{fidi}.
We have mapped the infinite interval of the $r$ coordinate $[0:\infty]$ to the
finite
compact interval $[0:1]$ using the new coordinate $\bar{z}:=r/(r+1)$.  
We have typically used grid sizes of $150$ points in $r$-direction and $70$
points in $\theta$ direction. The solutions pesented here have relative errors of $10^{-3}$
or smaller.

\section{$Q$-balls}
In this section, we study solutions in a flat space-time background, i.e.
we choose $G=0$. The $Q$-balls then interact only via the potential for $\lambda\neq 0$.

\subsection{Single $Q$-balls}
In order to be able to understand the structure of a system of two interacting $Q$-balls and their gravitating counterparts, 
we briefly reconsider
the one $Q$-ball system. This has been studied in great detail in \cite{vw,kk1,kk2,bh} and arises
from our model for $\lambda=0$ and $\phi_2\equiv 0$. 
In this section, we would like to emphasize some of the properties which are important to understand the
gravitating case.
In the present section on single $Q$-balls, we will omit the index $1$, respectively $2$ for all quantities. 
As was noticed in \cite{bh} -- using spherical coordinates $(r,\theta,\varphi)$ and a standard separation of variables -- the solutions to the linearized scalar field equation are given by:
\begin{equation}
\phi(r,\theta,\varphi)\propto \frac{J_{l+1/2}(\omega r)}{\sqrt{r}}Y_l^k(\theta,\varphi)
\end{equation}
where $J$ denotes the Bessel function and $Y_l^k$ 
are the standard spherical harmonics with $l$ integer and $-l \le k \le l$.
As was shown in \cite{bh}  -- at least for the first few values of $k$ and $l$ -- solutions to the full field equations exist in correspondence
to the symmetries of the spherical harmonics. Explicit examples of the angular excited solutions have  been presented for specific values of $k$ and $l$. In the following, these solutions will  be labelled
by the two quantum numbers $k$ and $l$, i.e. if we use $\phi_l^k$, we refer
to the solution of the full equation which corresponds to the spherical harmonic $Y_l^k$.

As already stated in Section 2.2, the boundary conditions depend on $l$, $k$. Here, we want to
state the boundary
conditions explicitely together with the parity $P$ of the solution 
under $\theta \rightarrow \pi-\theta$ for the first few values of $k$ and $l$ (see Table 1).
\begin{table}
\begin{center}
\begin{tabular}{|c|c|c|c|c|c|c|}  \hline
 l&k& $r=0$ & $r=\infty$ & $\theta = 0$ & $\theta = \pi/2$ & P  \\ \hline \hline
 0&0& $\partial_r\phi=0$&$\phi=0$&$\partial_{\theta}\phi=0$& $\partial_{\theta}\phi = 0$ & + \\ \hline
 1&0& $\partial_r\phi=0$&$\phi=0$&$\partial_{\theta}\phi=0$& $\phi = 0$  & - \\ \hline
1&1& $\phi=0$&$\phi=0$&$\phi=0$& $\partial_{\theta}\phi = 0$ & + \\ \hline 
2&0& $\partial_r\phi=0$&$\phi=0$&$\partial_{\theta}\phi=0$& $\partial_{\theta}\phi = 0$ 
& + \\ \hline
2&1& $\phi=0$&$\phi=0$&$\phi=0$& $\phi = 0$ 
& - \\ \hline
\end{tabular}
\caption{Boundary conditions and parity $P$ for different choices of $l$ and $k$}
\end{center}
\end{table}
Let us stress that all these conditions are compatible with the trivial solution $\phi\equiv 0$. 
Therefore, the numerical construction of non-vanishing solutions turns out to be a non trivial task. Solutions for the choices of $l$ and $k$ given in Table 1 have been presented in \cite{bh}. 

\subsection{Interacting $Q$-balls}
Interacting $Q$-balls were studied in detail in \cite{bh}. The interaction is characterized by
the coupling constant $\lambda$. The two $Q$-balls decouple in the limit $\lambda = 0$.
Here, we want to present the energy density of some of the solutions in order to illustrate the influence of the interaction term.

\begin{figure}[!htb]
\centering
\leavevmode\epsfxsize=11.0cm
\epsfbox{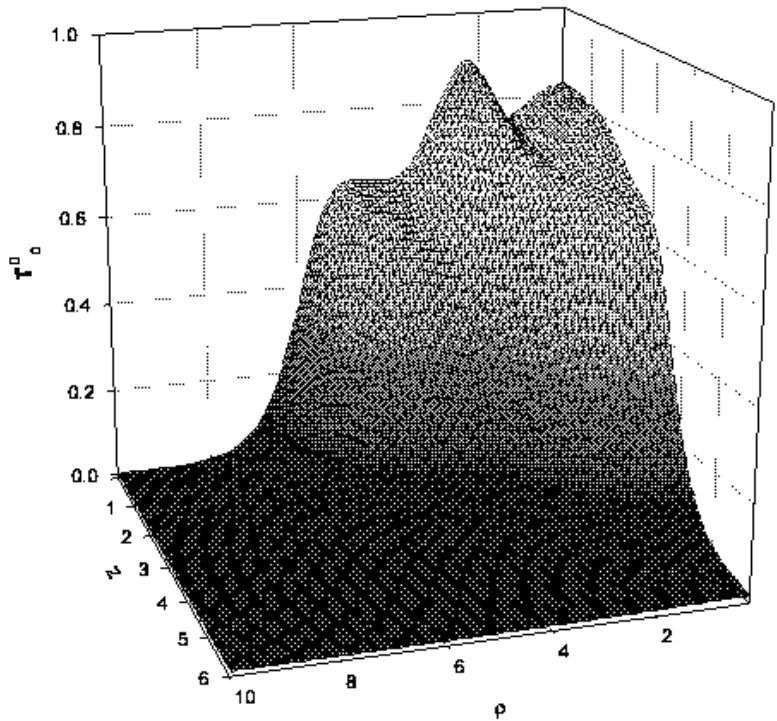}\\
\caption{\label{fig_mix_00_11} The energy density of two interacting $Q$-balls 
with $l_1=0$, $k_1=0$ (spherically symmetric, non-rotating) and $l_2=1$, $k_2=1$ (axially symmetric, rotating, parity even). Here $\lambda=1$ and $\omega_1=\omega_2=0.8$. Note that we use cylindrical coordinates $z=r\cos\theta$ and $\rho=r\sin\theta$.   }
\end{figure}

\begin{figure}[!htb]
\centering
\leavevmode\epsfxsize=13.0cm
\epsfbox{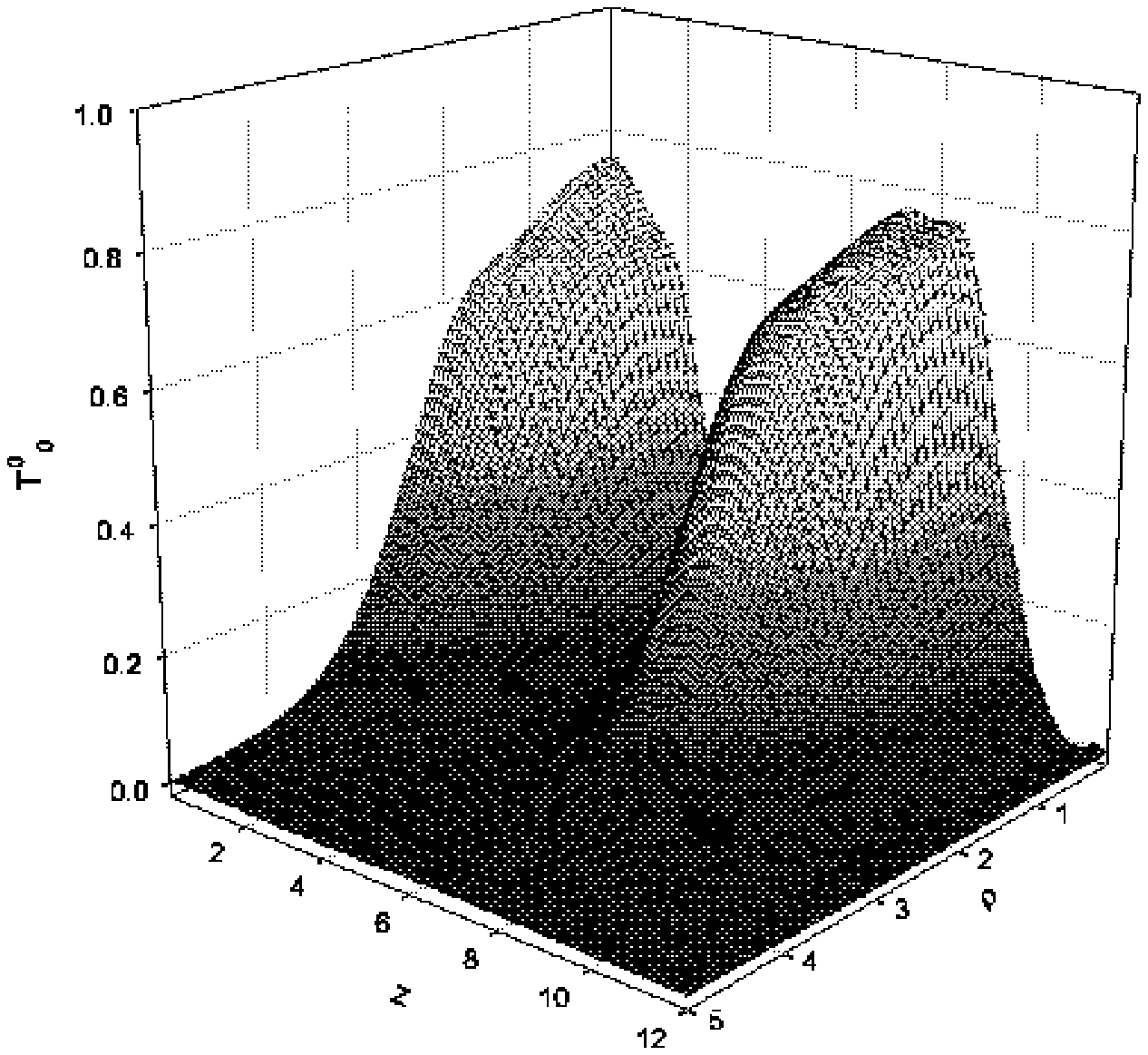}\\
\caption{\label{fig_mix_00_01} The energy density of two interacting $Q$-balls 
with $l_1=0$, $k_1=0$ (spherically symmetric, non-rotating) and $l_2=1$, $k_2=0$ (axially symmetric, non-rotating). Here $\lambda=1$ and $\omega_1=\omega_2=0.8$. Note that we use cylindrical coordinates $z=r\cos\theta$ and $\rho=r\sin\theta$.   }
\end{figure}

\begin{figure}[!htb]
\centering
\leavevmode\epsfxsize=13.0cm
\epsfbox{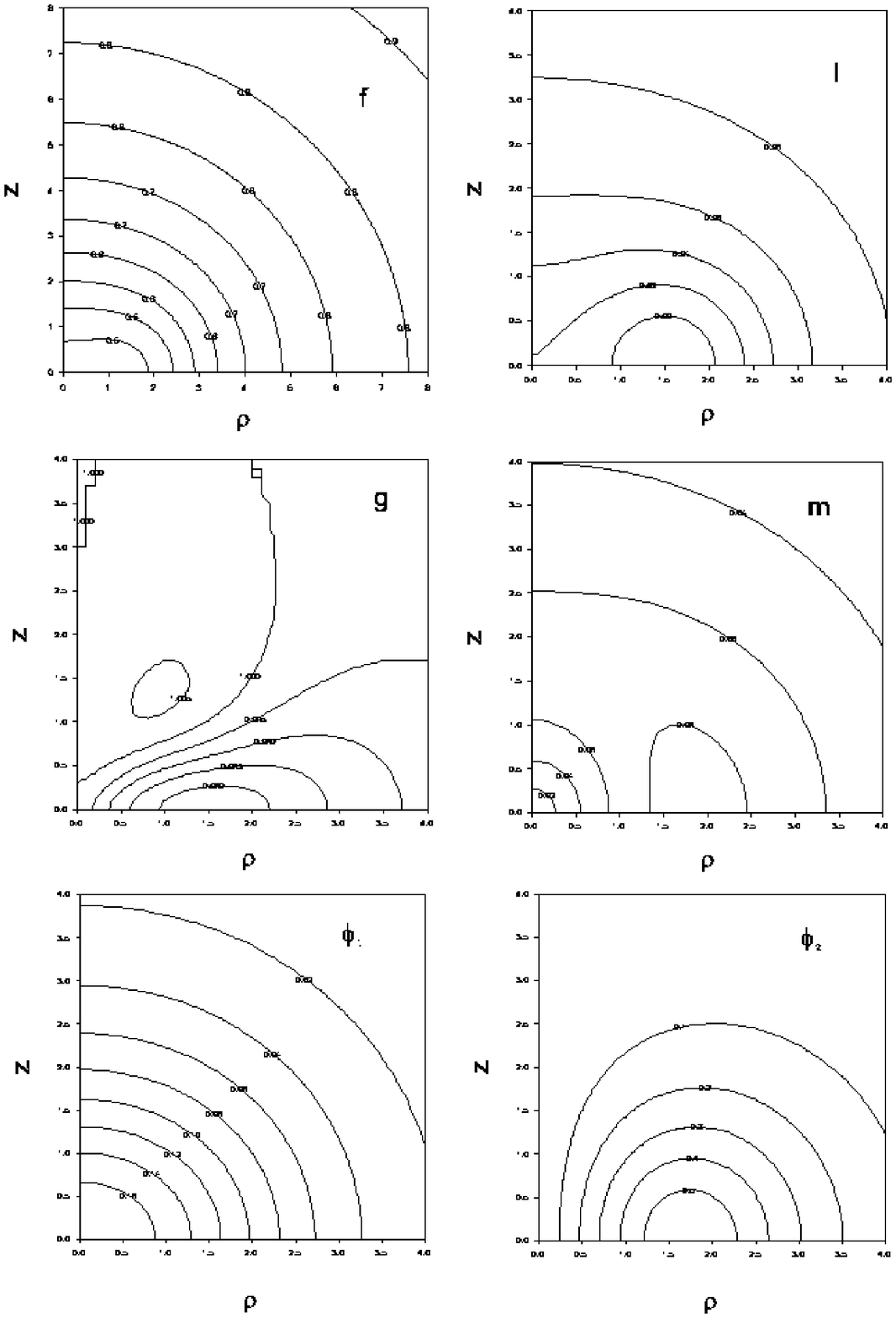}\\
\caption{\label{contour_00_11} The contour plots of the metric functions
$f$, $l$, $g$ and $m$ as well as of the scalar field functions
$\phi_1$ and $\phi_2$ are shown for two interacting boson stars
with $l_1=0$, $k_1=0$ (spherically symmetric, non-rotating) and $l_2=1$, $k_2=1$ (axially symmetric, rotating, parity even). Here $\alpha = 0.2$, $\lambda=0$ and $\omega_1=\omega_2=0.8$. Note that we use cylindrical coordinates $z=r\cos\theta$ and $\rho=r\sin\theta$.   }
\end{figure}

In Fig.\ref{fig_mix_00_11} we present the energy density for a spherically symmetric, non-rotating $Q$-ball interacting with
an axially symmetric, rotating and parity even $Q$-ball for $\lambda = 1$ and $\omega_1=\omega_2=0.8$. Note that we use cylindrical coordinates
here with $z=r\cos\theta$ and $\rho=r\sin\theta$. The local maximum appears at
$z=0$, $\rho \approx 2.2$ and is connected to the rotating $Q$-ball. In Fig. \ref{fig_mix_00_01} we give the energy density corresponding to a spherically symmetric, non-rotating $Q$-ball with $l_1=0$, $k_1=0$
interacting with an axially symmetric, non-rotating, i.e. angularly excited $Q$-ball with $l_2=1$, $k_2=0$ is given. Again, we have chosen $\lambda=1$ and $\omega_1=\omega_2=0.8$ for this case. The energy density has local maxima
at three different values of $z$. The energy density thus consists of two tori
at $z > 0$ and $z <0$ \footnote{Note that we plot the solution only for $z >0$, but that we 
can continue it to $z < 0$ due to its symmetry.} and a deformed spherical ball at $z=0$. 

\section{Boson stars}
Once gravity is included ($G\neq 0$), the different possible $Q$-ball solutions discussed in the
previous section become deformed by gravity and are called boson stars. 
Since only $k^2$ appears in the equation for the $Q$-ball the sign of $k$ does not matter and we can restrict ourselves to $k \ge 0$ without loosing generality in the flat space-time background case. This changes if we consider solutions in curved space-time, i.e. for $G > 0$. The reason is that the energy-momentum tensor involves terms that are linear in $k$.  
Reversing the sign of $k$ the solutions will have the same energy, but opposite
angular momentum.

In the case of interacting boson stars, the situation is more involved: when two boson stars  with $k_1$ and $k_2$ of the same sign interact the angular momentum is non-vanishing. However, for $k_1=-k_2$ the function $m(r,\theta)=0$ and the total angular momentum is vanishing.

In the presence of gravity, the boson stars interact by means of the direct coupling term in the potential for $\lambda\neq 0$ but also through gravity.
To understand the pattern of solutions we study the energy density in more detail. This reads
\be
\label{ed}
       T_0^0 = V(\phi_1,\phi_2) + \sum_{j=1}^2 
       \frac{f}{lg} (\partial_r \phi_j)^2 
     +  \frac{f}{r^2lg} (\partial_{\theta} \phi_j)^2
     + \frac{1}{f} \omega_j^2 \phi_j^2
     +   \left(\frac{f}{lr^2 \sin^2 \theta} - \frac{m^2}{f r^2} \right)k_j^2 \phi_j^2
\ee
The important point about this expression is that the term containing the
``rotation function'' $m(r,\theta)$ 
leads to a negative contribution to the energy density. Hence, rotation tends to decrease
the energy of the solution.

\subsection{Numerical results}

In the following, we will abbreviate 
$\alpha=8\pi G$ and focus on the case
$\omega_1 = \omega_2$.
We have fixed the parameters of the potential according to (\ref{parameters}).

\subsubsection{$l_1=0$, $k_1=0$, $l_2=0$, $k_2=0$}
This case describes two interacting non-rotating boson stars. The solutions are spherically symmetric. 
In particular, we find $g(r)=1$ and $m(r)=0$.
The bound state of two spherically
symmetric boson stars with a given  value of $\alpha$ behaves roughly as a single boson star with $\alpha/2$.
The mass $M$ and the values $\phi_i(0)$, $i=1,2$, $f(0)$, $l(0)$ and $T_0^0(0)$ slowly decrease when the parameter $\alpha$
increases. We found no evidence of a limiting behaviour, i.e. no critical
$\alpha$ beyond which the solutions cease to exist.

 \begin{figure}[!htb]
\centering
\leavevmode\epsfxsize=13.0cm
\epsfbox{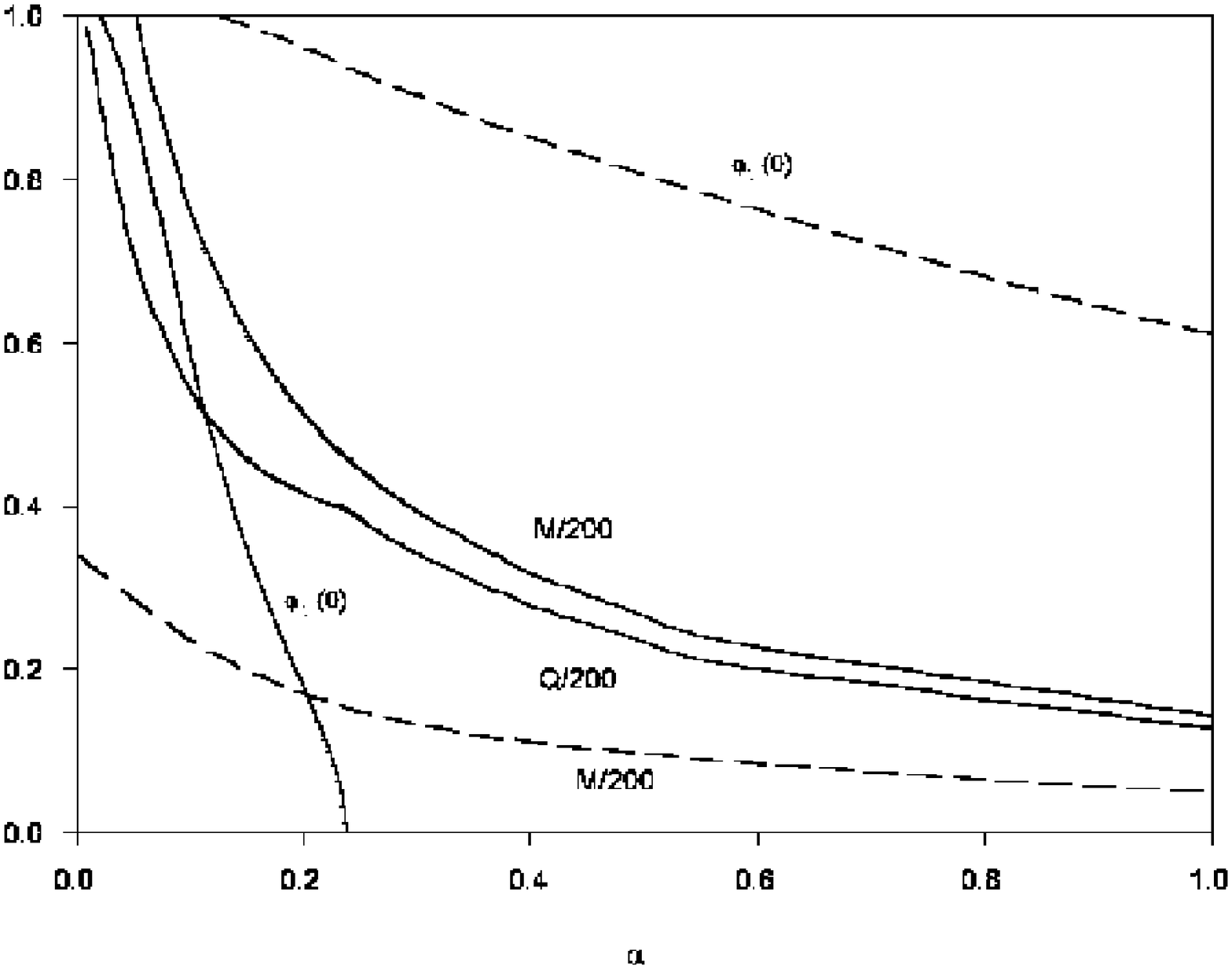}\\
\caption{\label{EJ_alpha_vary} The dependence of the mass $M$, charge $Q$ and $\phi_1(0)$ on $\alpha$ is shown for two interacting boson stars with
 $l_1=0$, $k_1=0$ (spherically symmetric, non-rotating) and $l_2=1$, $k_2=1$ (axially symmetric, rotating, parity even) (solid). Here $\lambda=0$. For comparison we also give the values for the $l_1=0$, $k_1=0$ single boson star for which
$\phi_2\equiv 0$ (dashed).}
\end{figure}

\begin{figure}[!htb]
\centering
\leavevmode\epsfxsize=13.0cm
\epsfbox{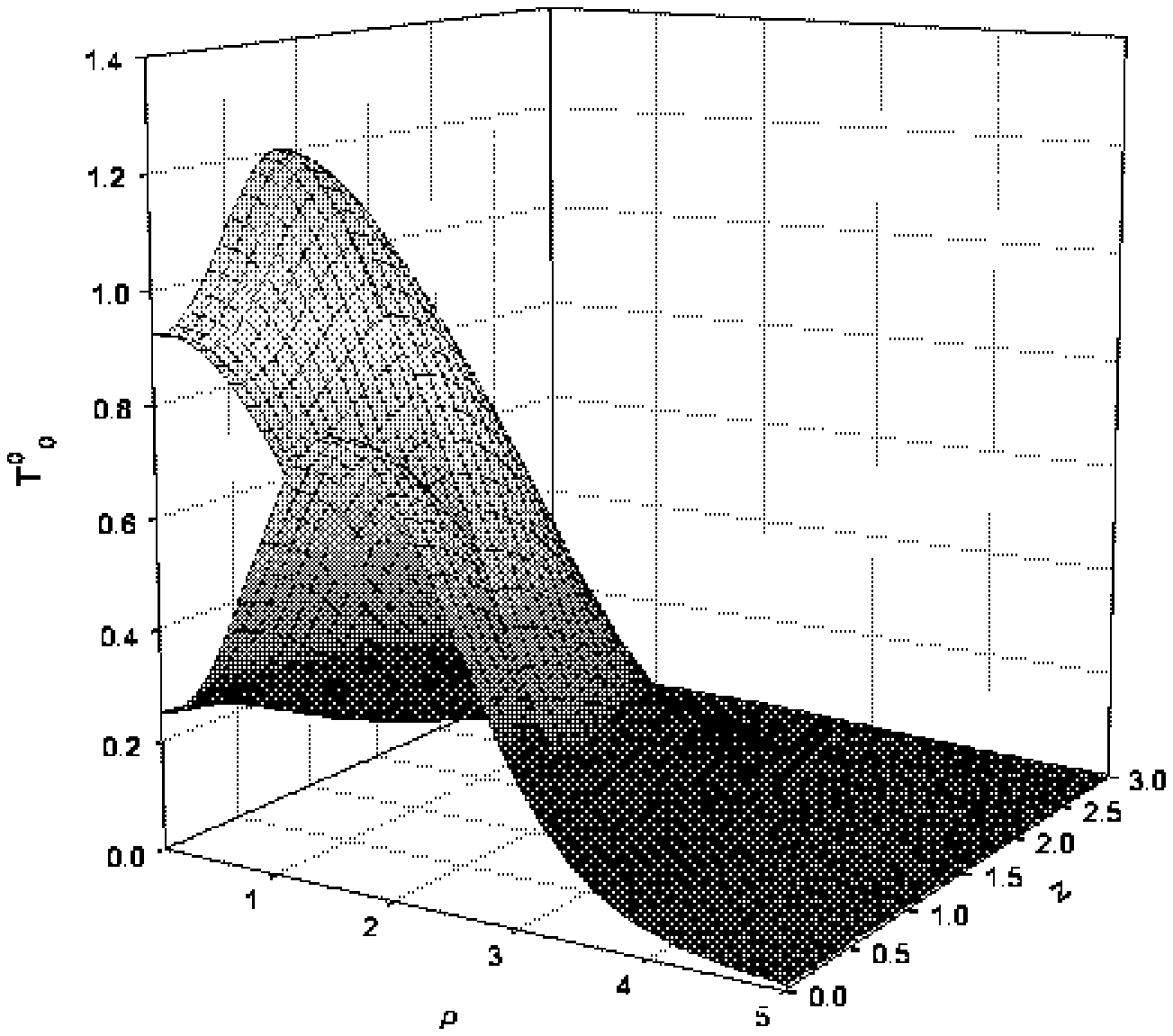}\\
\caption{\label{t00_00_11} The energy density of two interacting boson stars with
$l_1=0$, $k_1=0$ (spherically symmetric, non-rotating) and $l_2=1$, $k_2=1$ (axially symmetric, rotating, parity even) for $\alpha = 0.1$ (upper curve)
and $\alpha = 0.2$ (lower curve), respectively. Here $\lambda=0$ and $\omega_1=\omega_2=0.8$.
Note that we use cylindrical coordinates $z=r\cos\theta$ and $\rho=r\sin\theta$. }
\end{figure}

\subsubsection{$l_1=0$, $k_1=0$, $l_2=1$, $k_2=1$}
This corresponds to the interaction of a non-rotating boson star and a rotating, axially symmetric and parity even solution.

The contour plot of a typical solution of this type is presented in Fig. \ref{contour_00_11} for $\alpha = 0.2$, $\lambda=0$ and $\omega_1=\omega_2=0.8$.
Clearly, the metric functions show non-trivial behaviour now.

Varying the gravitational constant $\alpha$
(with all other parameters fixed), our numerical results show that the scalar field function $\phi_1$ of the spherically symmetric boson star
tends uniformly to zero and that only the spinning boson star survives for sufficiently large $\alpha$.
This result is consistent with the fact that rotation, i.e. non-vanishing
$m(r,\theta)$ tends to decrease the energy (see (\ref{ed})).
The rotating boson star is therefore energetically favoured in comparison
to the non-rotating case. 
This phenomenon is however strongly related to the gravitational interaction of the two boson stars. Indeed, the non-rotating
spherically symmetric boson alone (for $\phi_2\equiv 0$) exist for (arbitrarily) large values of $\alpha$. This is  
illustrated in Fig. \ref{EJ_alpha_vary} for $\lambda=0$ and $\omega_1=\omega_2=0.8$, where we plot
the mass $M$, charge $Q$ and $\phi_1(0)$, which corresponds to the maximal value of the
scalar field function $\phi_1(r)$ for non-rotating, spherically symmetric solutions
The quantities corresponding to interacting boson stars are given in solid lines. $\phi_1(0)$ and with this the field $\phi_1(r)$
corresponding to the non-rotating boson star
becomes identically zero for $\alpha= \alpha_{cr}\approx 0.25$ while the single boson star (dashed lines) exists for much larger values of $\alpha$. 
To illustrate this phenomenon further, we also plot 
the energy density of these solutions for two different values of $\alpha$ in Fig. \ref{t00_00_11}. The solution for $\alpha=0.1$ shows a quite large values for $T_0^0$ at 
the origin $\rho=0$, $z=0$. Since a non-vanishing contribution to the
energy density at the origin can only come from the spherically symmetric, non-rotating
boson star, the contribution of this boson star is still quite strong here.
When increasing $\alpha$, the axially symmetric, rotating boson star tends to absorb the
non-rotating one. This is clearly seen when comparing the $\alpha=0.1$ plot with that
for $\alpha=0.2$. Here, the value of $T_0^0$ at the origin
has decreased strongly and the energy density tends to the shape of a torus signalling that
the spherically symmetric contribution nearly vanishes.

We observe the qualitatively same effect for $\lambda\neq 0$. The critical value of the gravitational coupling $\alpha_{cr}$
where $\phi_1$ becomes identically zero depends slightly on the coupling constant $\lambda$. 
We find e.g. $\alpha_{cr}(\lambda=0.5) \approx 0.30$, 
$\alpha_{cr}(\lambda=0) \approx 0.25$ and
$\alpha_{cr}(\lambda=-0.5) \approx 0.22$, respectively.
 
Since the appearance of ergoregions for globally regular solutions signals the
existence of instabilities \cite{ergo} we have studies these here as well.
The ergoregions of single boson stars have been studied extensively in \cite{kk2}.
Ergoregions exist if the $g_{00}$ component of the metric becomes positive, i.e.
for
\begin{equation}
 g_{00}=-f+\frac{l m^2}{f} \sin^2 \theta  \ge 0  \ .
\end{equation}

\begin{figure}[!htb]
\centering
\leavevmode\epsfxsize=14.0cm
\epsfbox{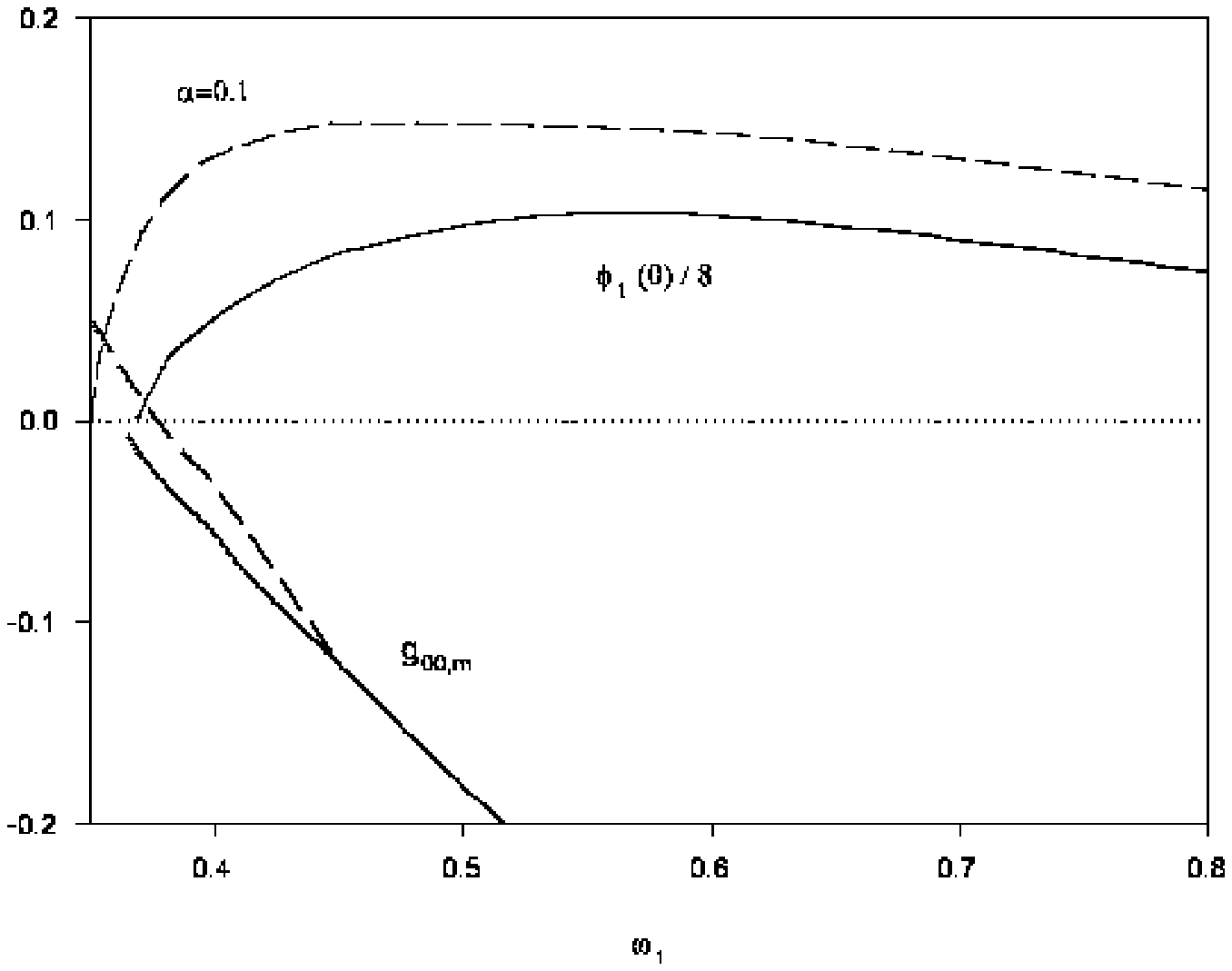}\\
\caption{\label{omega_00_11} The value of the scalar field function at the origin associated
to the spherically symmetric boson star $\phi_1(0)$
as well as the maximal value of the time component of the metric $g_{00,m}$ are
shown in dependence on $\omega_1=\omega_2$ for $\alpha=0.1$ and $\lambda=0$ (solid)
and $\lambda=1.5$ (dashed), respectively. Here $l_1=0$, $k_1=0$, $l_2=1$, $k_2=1$. }
\end{figure}

We have studied the appearance of ergoregions for $\alpha=0.1$, $\omega_1=\omega_2$ and two different values of $\lambda$, namely $\lambda=0$ and $\lambda=1.5$, respectively.
Our results are shown in Fig.\ref{omega_00_11}. Very similar to the case
for fixed $\omega_1=\omega_2$ and varying $\alpha$, the spherically symmetric, non-rotating
boson star disappears from the solution for $\omega_1 < \omega_{1,cr}$. We demonstrate this by plotting $\phi_1(0)$ ($\phi_1(r)$ has its maximal value at $r=0$) as function of $\omega_1$.
We find that $\omega_{1,cr}$ decreases with increasing $\lambda$, e.g. we find that
$\omega_{1,cr}\approx 0.37$ for $\lambda=0$ and $\omega_{1,cr}\approx 0.35$ for $\lambda=1.5$.
 This is easy to understand, since an increasing $\lambda$ means an increasing direct interaction between the two boson stars. We also plot the maximal value of $g_{00}$, $g_{00,m}$ as function of $\omega_1$.
Interestingly, for $\lambda=0$ $g_{00,m}$ stays negative as long as the spherically symmetric
boson star is present. Thus for $\alpha=0.1$ and $\lambda=0$, there are no instabilities
due to ergoregions for those values of $\omega_1=\omega_2$ where genuine interacting boson
stars exist. For smaller values of $\omega_1=\omega_2$ (and the second branch of solutions), the curve for $g_{00,m}$ would
follow the curve for a single rotating boson star given in \cite{kk2}.
The situation changes for $\lambda =1.5$. Here, $g_{00,m}$ becomes zero at a value of
$\omega_1=\omega_1^{(0)}$ which is larger that the value $\omega_{1,cr}$ at which the 
spherical boson star disappears from the system. We find $ \omega_1^{(0)}\approx 0.38$ for $\kappa=0.1$ and $\lambda=1.5$. Thus for $\omega_{1,cr} \le \omega_1 \le \omega_1^{(0)}$ the two interacting boson stars possess an ergoregion signalling an instability. It appears that the stronger direct interaction between the boson
stars tends to destabilize the system. A more detailed investigation
and plots of the ergoregions will be presented in a future publication.

\subsubsection{$l_1=0$, $k_1=0$, $l_2=2$, $k_2=1$}

\begin{figure}[!htb]
\centering
\leavevmode\epsfxsize=11.0cm
\epsfbox{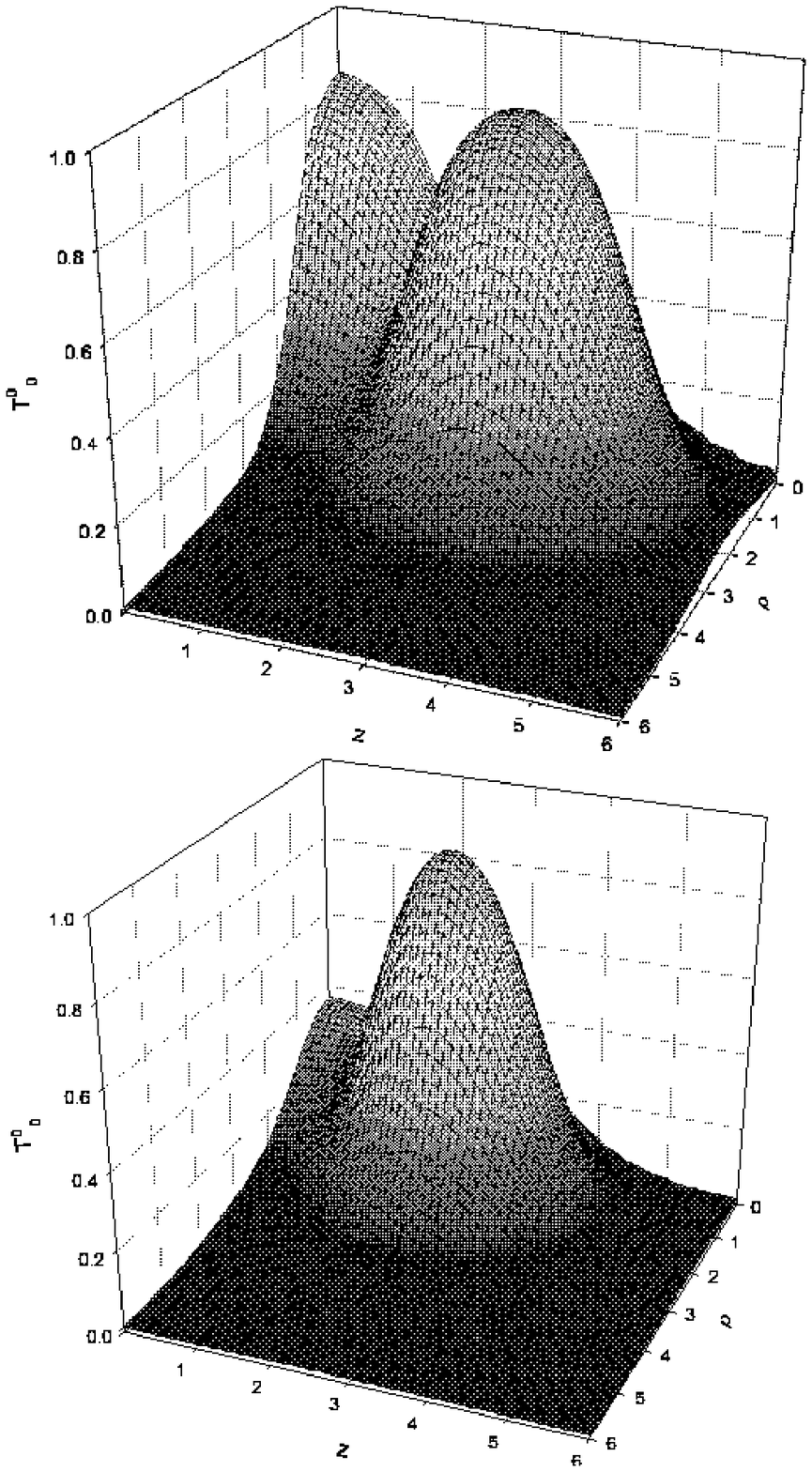}\\
\caption{\label{t00_00_21} The energy density of two interacting boson stars with
$l_1=0$, $k_1=0$ (spherically symmetric, non-rotating) and $l_2=2$, $k_2=1$ (axially symmetric, rotating, parity odd). Here $\lambda=0$, $\omega_1=\omega_2=0.8$ and $\alpha = 0.05$ (top)
and $\alpha = 0.2$ (bottom), respectively. Note that we use cylindrical coordinates $z=r\cos\theta$ and $\rho=r\sin\theta$.  }
\end{figure}

\begin{figure}[!htb]
\centering
\leavevmode\epsfxsize=13.0cm
\epsfbox{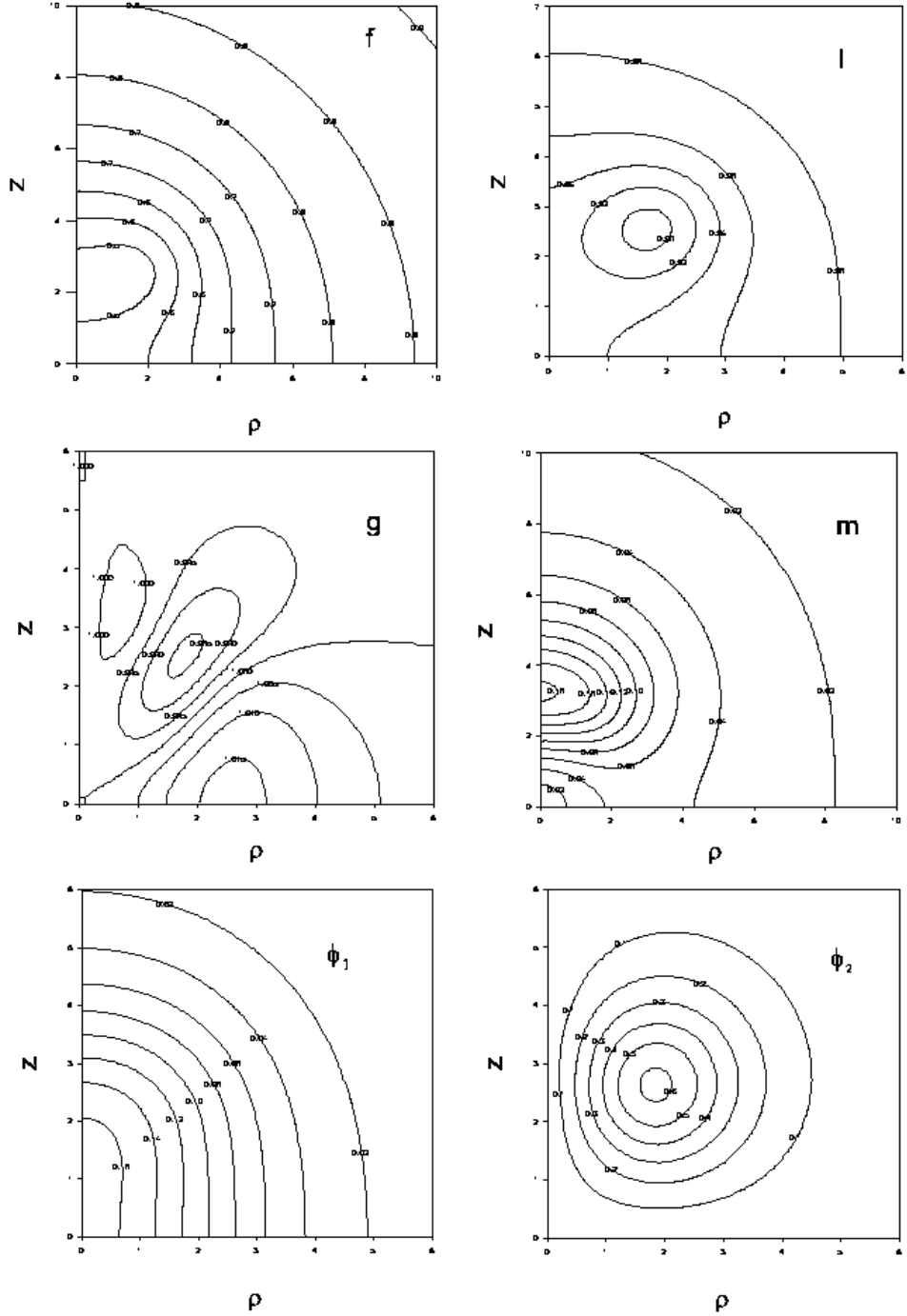}\\
\caption{\label{fignew} The contour plots of the metric functions
$f$, $l$, $g$ and $m$ as well as of the scalar field functions
$\phi_1$ and $\phi_2$ are shown for two interacting boson stars
with $l_1=0$, $k_1=0$ (spherically symmetric, non-rotating) and $l_2=2$ and $k_2=1$ (axially symmetric, rotating, parity odd). Here $\alpha=0.1$, $\lambda=0$ and $\omega_1=\omega_2=0.8$.   }
\end{figure}

This corresponds to the interaction of a non-rotating boson star and a rotating, axially symmetric and parity odd -- or in our notation ``angularly excited'' -- solution.
The $\phi_2$ component corresponding to the rotating boson star is odd under the $z \to -z$ reflexion
Our numerical results indicate that this case is qualitatively similar to the previous case.
  For vanishing and small $\alpha$  
 the energy density  has its  maximum in a ball centered around the origin as well as  
 in two tori at $z >0$ and $z <0$. This is clearly seen in Fig.\ref{t00_00_21}, where
we plot the energy density of the solutions for two different values of $\alpha$, $\lambda=0$ 
and $\omega_1=\omega_2=0.8$.
When the parameter $\alpha$ increases, 
the spherically symmetric boson star disappears and the solution becomes purely axially symmetric. This is indicated in Fig.\ref{t00_00_21} for $\alpha=0.2$ and $\alpha=0.05$.
Again for $\alpha=0.05$, the energy density has a big value at the origin which results
from the non-rotating boson star. For $\alpha=0.2$ this value has decreased significantly
and the energy density looks like that of an axially symmetric solution.
We also present the contour plots of the metric and matter field functions in Fig.\ref{fignew} 
for $\alpha=0.1$, $\lambda=0$ and $\omega_1=\omega_2=0.8$.

\begin{figure}[!htb]
\centering
\leavevmode\epsfxsize=14.0cm
\epsfbox{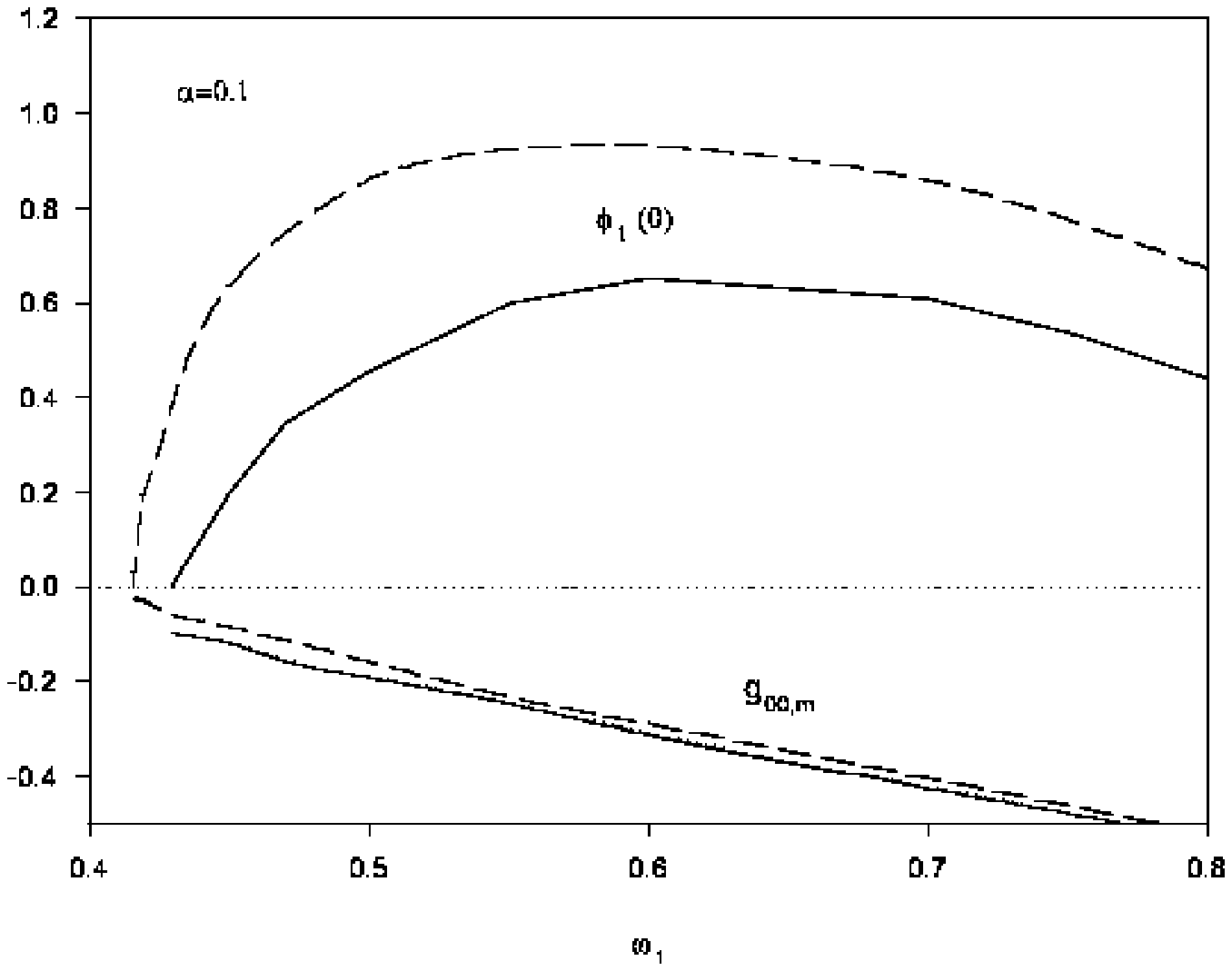}\\
\caption{\label{omega_00_21} The value of the scalar field function at the origin associated
to the spherically symmetric boson star $\phi_1(0)$
as well as the maximal value of the time component of the metric $g_{00,m}$ are
shown in dependence on $\omega_1=\omega_2$ for $\alpha=0.1$ and $\lambda=0$ (solid)
and $\lambda=1.5$ (dashed), respectively. Here $l_1=0$, $k_1=0$, $l_2=2$, $k_2=1$.   }
\end{figure}

Considering the ergoregions of these solutions
we find that for $\alpha=0.1$ and both values of $\lambda=0$ and $\lambda=1.5$, respectively,
$g_{00,m}$ stays negative for all values of $\omega_1=\omega_2$ for which genuine
interacting boson stars exist. When interacting with a rotating, parity odd boson star the spherically symmetric boson star disappears from the solution for values of $\omega_{1}$ larger than in the case of interaction with a parity even boson star (see previous section).
For $\omega_1=\omega_2$ smaller than $\omega_{1,cr}$, i.e. the frequency at which $\phi_1(0)$ (the maximal value of $\phi_1(r)$) vanishes (and for the second branch of solutions) the curve follows
thus the curve of a rotating, parity odd boson star presented in \cite{kk2}.
We find that $\omega_{1,cr}\approx 0.43$ for $\kappa=0.1$ and $\lambda=0$, while
$\omega_{1,cr}\approx 0.41$ for $\kappa=0.1$ and $\lambda=1.5$.

\section{Conclusion}
In this paper we have studied angularly excited as well as interacting $Q$-balls 
and their gravitating counterparts, boson stars. When 
a non-rotating boson star interacts with a rotating boson star
for fixed angular frequency $\omega$ and varying gravitational coupling the non-rotating
boson star tends to disappear from the system if the gravitational coupling reaches a critical
value. For gravitational couplings above these critical values, only single, axially symmetric
boson stars exist. The same holds true if the gravitational coupling is fixed
and the angular frequency lowered. Then below a critical value of the angular frequency $\omega$ the non-rotating boson stars has disappeared from the bound system. As a consequence the bound state
of two boson stars with $l_1=k_1=0$ and $l_2=k_2=1$ exists only on a finite domain of the parameter
space $\omega_1$, $\omega_2$.
We observe that this behaviour is qualitatively independent of the
choice of the direct interaction parameter $\lambda$ in the potential.

When considering the existence of ergoregions for our solutions, which would
signal an instability, we observe that the two cases of a non-rotating
boson star interacting with a rotating, parity even boson star on the one hand
and with a rotating, parity odd boson star on the other hand, are qualitatively different.
While in the latter case no ergoregions appear for genuinely interacting boson
stars, i.e. with the non-rotating boson star present in the system, ergoregions
appear in the former case if the interaction parameter $\lambda$ is made large enough.


\begin{thebibliography}{99}
\bibitem{ms} N.S. Manton and P.M. Sutcliffe, 
{\it Topological solitons}, Cambridge University Press, 2004.
\bibitem{fls} R. Friedberg, T. D. Lee and A. Sirlin, Phys. Rev. D {\bf 13} (1976) 2739.
\bibitem{lp} T. D. Lee and Y. Pang, Phys. Rep. {\bf 221} (1992), 251.
\bibitem{coleman}  S. R. Coleman, Nucl. Phys. B {\bf 262} (1985), 263.
\bibitem{vw} M.S. Volkov and E. W\"ohnert, Phys. Rev. D {\bf 66} (2002), 085003.
\bibitem{kusenko} A. Kusenko, Phys. Lett. B {\bf 404} (1997), 285; Phys. Lett. B {\bf 405} (1997), 108.
\bibitem{dm} {\it see e.g.} A. Kusenko, hep-ph/0009089.
\bibitem{implications} K.~Enqvist and J.~McDonald, Phys.\ Lett.\ B {\bf 425} (1998), 309;
S.~Kasuya and M.~Kawasaki, Phys.\ Rev.\  D {\bf 61} (2000), 041301;
A.~Kusenko and P.~J.~Steinhardt, Phys.\ Rev.\ Lett.\  {\bf 87} (2001), 141301;
T.~Multamaki and I.~Vilja, Phys.\ Lett.\  B {\bf 535} (2002), 170;
M.~Fujii and K.~Hamaguchi, Phys.\ Lett.\  B {\bf 525} (2002), 143;
M.~Postma, Phys.\ Rev.\  D {\bf 65} (2002), 085035;
K.~Enqvist, {\it et al.}, Phys.\ Lett.\  B {\bf 526} (2002), 9;
M.~Kawasaki, F.~Takahashi and M.~Yamaguchi, Phys.\ Rev.\  D {\bf 66} (2002), 043516;
A.~Kusenko, L.~Loveridge and M.~Shaposhnikov,
Phys.\ Rev.\  D {\bf 72} (2005), 025015;
Y.~Takenaga {\it et al.}  [Super-Kamiokande Collaboration],
Phys.\ Lett.\  B {\bf 647} (2007), 18; S.~Kasuya and F.~Takahashi, JCAP {\bf 11} (2007), 019.
\bibitem{cr} L. Campanelli and M. Ruggieri, Phys. Rev. D {\bf 77} (2008), 043504. 
\bibitem{kk1} B. Kleihaus, J. Kunz and M. List, Phys. Rev. D {\bf 72} (2005), 064002.
\bibitem{kk2} B. Kleihaus, J. Kunz, M. List and I. Schaffer, Phys. Rev. D {\bf 77} (2008), 064025.
\bibitem{bh} Y. Brihaye and B. Hartmann, Nonlinearity {\bf 21} (2008), 1937. 
\bibitem{radu} E. Radu, {\it On rotating solutions in General Relativity}, arXiv: gr-qc/0512094.
\bibitem{nonspherical} 
S.~Kasuya and M.~Kawasaki, Phys. Rev. D {\bf 62} (2000) 023512; Phys. Rev. Lett. {\bf 85} (2000) 2677;
Phys. Rev. D. {\bf 64} (2001) 123515; T. Multamaki and I. Vilja, Nucl. Phys. B {\bf 574} (2000), 130; K. Enqvist, A. Jokinen, T. Multamaki and I. Vilja, Phys. Rev. D {\bf 63} (2001), 083501.
\bibitem{km2} A. Kusenko and A. Mazumdar, Phys. Rev. Lett. {\bf 101} (2008), 211301.
\bibitem{misch} E. Mielke and F. E. Schunck, Proc. 8th Marcel Grossmann Meeting, Jerusalem, Israel, 22-27
Jun 1997, World Scientific (1999), 1607.
\bibitem{flp} R. Friedberg, T. D. Lee and Y. Pang, Phys. Rev. D {\bf 35} (1987), 3658.
\bibitem{jetzler} P. Jetzer, Phys. Rept. {\bf 220} (1992), 163.
\bibitem{fidi} W. Sch\"onauer and R. Wei\ss, J. Comput. Appl. Math. {\bf 27} (1989) 279;
M. Schauder, R. Wei\ss and W. Sch\"onauer, ``The CADSOL Program Package'', Universit\"at Karlsruhe, Interner
Bericht Nr. 46/92 (1992); W. Sch\"onauer and E. Schnepf, ACM
Trans. Math. Softw. {\bf 13} (1987) 333.
\bibitem{ergo} V. Cardoso, P. Pani, M. Cadoni and M. Cavaglia, Phys. Rev. D {\bf 77} (2008), 124044; J. L. Friedman, Commun. Math. Phys. {\bf 63} (1978), 243;
N. Comins and B. F. Schutz, Proc. R. Soc. Lond. A {\bf 364} (1978), 211;
S. Yoshida and Y. Eriguchi, MNRAS {\bf 282} (1996), 580. 
\end{thebibliography}
\end{document}